\documentclass[twocolumn,showpacs,preprintnumbers,amsmath,amssymb]{revtex4}

\usepackage{amsmath}
\usepackage{amsfonts}
\usepackage{amssymb, multirow}
\usepackage{pdfsync}
\usepackage{epsfig}
\usepackage{graphicx}
\usepackage{subfigure}
\usepackage{nicefrac}

\newcommand{\be}{\begin{equation}}\newcommand{\ee}{\end{equation}}
\newcommand{\bea}{\begin{eqnarray}} \newcommand{\eea}{\end{eqnarray}}
\newcommand{\ba}[1]{\begin{array}{#1}} \newcommand{\ea}{\end{array}}

\long\def\symbolfootnote[#1]#2{\begingroup%
\def\thefootnote{\fnsymbol{footnote}}\footnote[#1]{#2}\endgroup} 

\def\p{\partial }

\def\rr{{\rm r}}

\newcommand{\cK}{{\cal K}}

\newcommand{\Kahler}{K{\" a}hler~}

\newcommand{\RN}{Reissner-Nordstr{\" o}m~}

\def\p{\partial}
\def\bfone{\relax{\rm 1\kern-.35em 1}}

\begin{document}

\preprint{ICCUB-12-223}

\title{Superconductors for superstrings on $AdS_5\times T^{1,1}$}

\author{Francesco Aprile$^1$, Andrea Borghese$^2$, Aldo Dector$^1$, 
 Diederik Roest$^2$ and Jorge G. Russo$^{1,3,4}$}
\affiliation{%
~ \\
1) Institute of Cosmos Sciences and ECM 
- Universitat de Barcelona, Av. Diagonal 647,  08028 Barcelona, Spain \\
2) Centre for Theoretical Physics - University of Groningen, Nijenborgh 4 9747 AG Groningen, The Netherlands \\
3)   Perimeter Institute for Theoretical Physics, Waterloo, Ontario, N2L 2Y5, Canada \\
4) Instituci\'o Catalana de Recerca i Estudis Avan\c cats (ICREA),
Pg. Lluis Companys, 23, 08010 Barcelona, Spain
}%

\begin{abstract}

We construct a one-parameter family of five-dimensional ${\cal N}=2$
supergravity Lagrangians with an $SU(2,1)/U(2)$ hypermultiplet. For
certain values of the parameter, these are argued to describe the
dynamics of scalar modes of superstrings on $AdS_5\times T^{1,1}$, 
and therefore to be dual to specific chiral primary operators of
Klebanov-Witten superconformal field theory. We demonstrate that,
below a critical temperature, the thermodynamics is dominated by
charged black holes with hair for the scalars that are dual to the
operator of lowest conformal dimension $\nicefrac32$. The system thus
enters into a superconducting phase where $\langle {\rm Tr}[A_kB_l]
\rangle$ condenses.

\end{abstract}

\maketitle


Holographic superconductors provide an interesting laboratory
to investigate applications of the AdS/CFT correspondence to condensed matter \cite{Gubser:2008px, Hartnoll:2008kx}.
The prototypical example of such a correspondence is based on string compactification on $AdS_5\times S^5$. In this case the field theory description for D3-branes in flat space-time is in terms of 
${\cal N}=4$ Super Yang-Mills theory with chemical potentials.
The dual gravitational system is described by five-dimensional ${\cal N}=8$ $SO(6)$ gauged supergravity, whose holographic superconducting phases were investigated in \cite{Aprile:2011uq}. On both sides of the correspondence, the field-theory dynamics is tightly constrained
by the high degree of supersymmetry.
It would clearly be desirable to have new setups for  superconductors with less supersymmetry
and a well-understood  dual field theory.

A class of ${\cal N}=1$ superconformal quiver gauge theories is described by type IIB compactifications on Sasaki-Einstein  manifolds.
An example  known in great detail is the Klebanov-Witten (KW) superconformal theory for D3-branes on the conifold \cite{Klebanov:1998hh}. The dual description involves superstrings on  $AdS_5\times T^{1,1}$. The latter factor is the five-dimensional coset space ${SU(2)\times SU(2)}/{U(1)}$ with a diagonal embedding of the denominator in the numerator \cite{Romans:1984an}. 
The compactification preserves 8 supersymmetries, in agreement with the number of supersymmetries of an 
${\cal N}=1$ superconformal field theory.

The KW theory is an $SU(N)\times SU(N)$ gauge
theory with chiral superfields $A_k$, $B_l$, $k,l = 1, 2$,  transforming in the 
$(N,\bar N)$ and  $(\bar N,N)$ bi-fundamental representations, respectively, and forming doublets
of the $SU(2)\times SU(2)$ global symmetry group.
In the non-trivial IR fixed point, the fields have conformal dimension $3/4$, in such a way that the 
 quartic superpotential
$W = \lambda {\rm Tr}(A_1 B_1 A_2 B_2 - A_1 B_2 A_2 B_1)$ becomes marginal.

In the context of holographic superconductivity, the low-temperature
thermodynamics is typically dominated by the chiral operators of lowest conformal dimension.
In  KW theory, this operator is 
\be
O_{k,l}={\rm Tr} (A_k B_l) \ . \label{operator}
\ee
It has $R$-charge equal to 1 and conformal dimension $\nicefrac32$. The aim of this paper is to construct a  gravity system
that describes the dynamics of its  dual fields and to investigate its holographic superconducting phases.

The relevant Lagrangian must be described by ${\cal N}=2$ gauged supergravity coupled
to at least one hypermultiplet, necessary to describe the degrees of freedom of this operator.
Here we will focus on a particular model with  one hypermultiplet,
whose scalar manifold is given by $SU(2,1) / U(2)$.
This is the unique homogeneous 4-manifold that is both K{\" a}hler and 
quaternionic-K{\" a}hler. We will not consider non-homogeneous manifolds as these are not expected to emerge in reductions over highly symmetric spaces
 such as $T^{1,1}$.
We denote by $\zeta_i=\{ \zeta_+, \zeta_- \}$ the two complex scalar fields of the hypermultiplet. Their kinetic terms follow from the \Kahler potential $\cK = - \ln(1-|\zeta_+|^{2}-|\zeta_-|^{2})$. The graviphoton will gauge a compact isometry of $SU(2)\times U(1)$, whose components in the most general form read
\begin{eqnarray}
K^{\zeta_{\pm}}=-\tfrac12 \sqrt{6} i  \left(\left(\beta \pm \alpha_{3}\right)\zeta_{\pm}+\left(\alpha_{1} \mp i\alpha_{2}\right)\zeta_{\mp}\right)\,, \notag
\end{eqnarray}
and $K^{\bar{\zeta}_{i}}=(K^{\zeta_i})^{\star}$. Due to the SU(2) invariance of the model, in what follows we will set $\alpha_{1}=\alpha_{2}=0$ without loss of generality. Furthermore, we will fix the overall scale by setting $\alpha_3 = 1$.  The remaining parameter $\beta$ specifies the embedding of the Abelian gauge group and leads to a one-parameter family of theories. 

The bosonic part of the Lagrangian describing the coupling between the supergravity and the hypermultiplet can be constructed from the formulas in \cite{Ceresole:2001wi}. We find:
\bea
 2\kappa^2 e^{-1}\mathcal{L} &= & R-\frac{1}{4}F_{\mu\nu}F^{\mu\nu} + \frac{1}{12\sqrt{3}}\epsilon^{\mu\nu\rho\sigma\tau} A_\mu F_{\nu\rho}F_{\sigma\tau} 
 + \notag \\
  &&  - 2 \frac{\delta^2 \cK}{\delta \zeta^i  \, \delta \zeta^{\bar j}} D_{\mu}\zeta^{i}D^{\mu}\bar{\zeta}^{\bar j} - \mathcal{V} \label{LL}
\,,\notag \\
D_{\mu}\zeta_{\pm}&=& \partial_{\mu}\zeta_{\pm} - \tfrac 12 \sqrt{3}  i  \rr_\pm  A_{\mu}\zeta_\pm \,,  
 \eea
with the $R$-charges given by $\rr_{\pm}=  \beta \pm 1$ and
\begin{eqnarray}
\mathcal{V} &=& - \frac{3}{2(1-|\zeta_{+}|^2-|\zeta_{-}|^2)^2} \Big(\ 8+2(1-\beta )^{2}|\zeta_{+}|^{4}+ \notag \\
 && + 2(1+\beta )^{2}|\zeta_{-}|^{4}+4(2+\beta ^{2})|\zeta_{+}|^2|\zeta_{-}|^2 + \nonumber\\
 &&-(11-2\beta +3\beta ^{2})|\zeta_{+}|^{2}-(11+2\beta +3\beta ^{2})|\zeta_{-}|^{2}\ \Big)\ . \notag
\end{eqnarray}
At the origin $\mathcal{V}(0)=-12$, yielding an $AdS_5$ solution of unit radius. In this critical point, the masses of $\zeta_\pm$ are
\be
m_{\pm}^{2}=- \tfrac{3}{4} \ (\beta \pm 1)(5 \mp 3\beta) \,. \notag
\ee
The Lagrangian is  symmetric under $\zeta_{+}\leftrightarrow\zeta_{-}$ with  $\beta \leftrightarrow-\beta$.
Therefore without loss of generality we can take  $\beta \geq 0 $. Table 1 summarizes the different hypermultiplets described by our
 Lagrangian for integer $\beta$ up to $3$. We also indicate the conformal dimension $\Delta$,  related to the mass via the standard AdS/CFT relation $m^2 = \Delta (\Delta - 4)$.

\begin{table}[ht]
\begin{center}
\begin{tabular}{|c||c|c|c|}
\hline
		$\beta$   		& $m_\pm^2$ 			 & $\rr_\pm $ & $\Delta_\pm$ \\ 
\hline \hline
	 $0$ &  $(-\nicefrac{15}{4},-\nicefrac{15}{4})$   			 &  $(1,-1)$  & $(\nicefrac{3}{2},\nicefrac{5}{2})$ \\
\hline
 $1$ &  $(-3,0)$   			 &  $(2,0)$   & $(3,4)$   \\
\hline
 $2$ &  $(\nicefrac{9}{4},\nicefrac{33}{4})$   			 &  $(3,1)$  & $(\nicefrac{9}{2},\nicefrac{11}{2})$ \\
\hline
 $3$ & $(12,21)$ 			 &  $(4,2)$  & $(6,7)$ \\
\hline
\end{tabular}	
\end{center}
\caption{\it The mass, charge and conformal dimension of the complex scalars in the hypermultiplet for various values of $\beta$.}
\end{table}

\begin{figure}[ht]
\centering
\includegraphics[scale=.5]{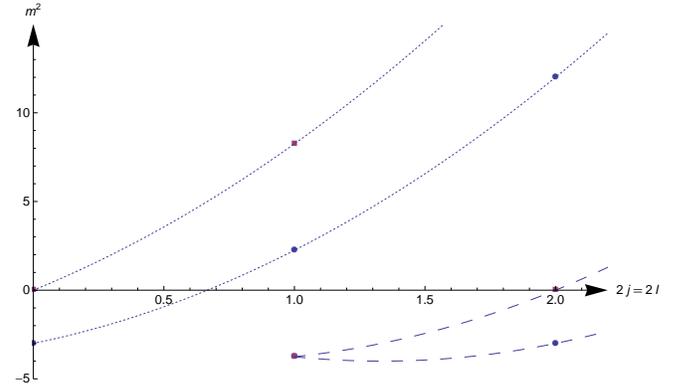}
\caption{\it \small{The  two KK towers of scalar fields on $T^{1,1}$. The dotted (dashed) line indicates hypers with $2j = \beta \mp 1$.
}}
\label{fig:KK}
\end{figure}

For $\beta$ integer, the masses and charges are in precise correspondence with those of chiral $AdS$ multiplets for type IIB on $AdS_5\times T^{1,1}$, with specific $SU(2) \times SU(2)$ quantum numbers $(j,l)$. In particular, one has the following two Kaluza-Klein (KK) towers \cite{Ceresole:1999zs, Baumann:2010sx}:
 \begin{itemize}
 \item
 The complex IIB zero- and two-forms give rise to a KK tower with $2j = 2l= \beta - 1$ with $\beta \geq 1$. It is dual to the field theory operators
 \be
  {\rm Tr} [(W_1^2 + W_2^2)(A_k B_l)^{\beta -1}] + \ldots \,. \label{tower1}
 \ee
\item
A second KK tower originates from the IIB metric, four-form, complex two-forms and has $2j=2l= \beta + 1$ with $\beta \geq 0$.  The corresponding operators are 
  \be
  {\rm Tr} [(A_k B_l)^{\beta +1}] \,. \label{tower-operator}
 \ee
 \end{itemize}
In addition there is also an isolated Betti hypermultiplet with $2j=2l=0$ and $\beta=1$,  dual to the operator ${\rm Tr} [W_1^2 - W_2^2]$. The resulting mass spectrum as a function of $2j=2l$ is shown in fig.~\ref{fig:KK}. In the following we discuss the lowest components of the two KK towers.
These correspond to the most interesting operators for the thermodynamical study of the theory at finite density. 

The lowest KK state in the first tower has $\beta =1$  with  $ m^2= -3$ and $m^2=0$, and is often referred to as the universal hypermultiplet.  In this case the additional terms in \eqref{tower1} correspond to the superpotential and Kahler potential of conformal dimensions $\Delta = 3$ and $4$, respectively. Indeed $\zeta_+$ is nothing but the complex scalar field that appears in the truncation by Gubser et al \cite{Gubser:2009qm}. Furthermore, the massless scalar $\zeta_-$ originates from  the  type IIB complex dilaton 
\cite{Cassani:2010uw, Liu:2010sa, Gauntlett:2010vu, Skenderis:2010vz}. Explicitly, the field redefinition
\bea
&&\zeta_+=\sqrt{1 - |\zeta_-|^2}\ \tanh {\tfrac12 \eta}\ e^{i\varphi}\ \sqrt{\frac{1+i\bar\tau}{1-i\tau}} \,, \notag \\
&&\zeta_- = \frac{1 + i\tau}{1 - i\tau}\ ,\qquad \tau =C_0+ie^{-\phi }\ , \notag
\eea
takes the second line of (\ref{LL}) into the form given in  \cite{Cassani:2010uw, Liu:2010sa, Gauntlett:2010vu}:
\bea
&& - \tfrac12 (\p \eta)^2  - \tfrac12  e^\phi \sinh^2\eta \ \p^\mu C_0 \big(\p_\mu \varphi - \sqrt{3} A_\mu   \big)
 \notag \\
&& - \tfrac12  \sinh^2\eta \ (\p \varphi - \sqrt{3} A )^2  - \tfrac12 e^{2\phi}\cosh^4(\tfrac12 \eta) 
(\p C_0)^2
 \notag \\
&& -  \tfrac12 \cosh^2(\tfrac12 \eta)\ (\p\phi)^2 - 3 \cosh^2 (\tfrac12 \eta) \big( \cosh\eta -\ 5\big) \ . \qquad
\label{dallag}
\eea
The KK states in the first tower  with $\beta \geq 2$ have positive mass square and higher conformal dimensions and are not relevant for the thermodynamics describing $U(1)_R$ spontaneous symmetry breaking (SSB), so they will not be considered here.

The lowest KK state in the second tower has $\beta =0$ and hence two conformally coupled complex scalars with identical masses $ m^2 = -\nicefrac{15}{4}$. These are dual to a chiral operator of conformal dimension $\nicefrac32$. There is a unique operator in KW theory with this  dimension, given by the lowest mode \eqref{operator} of the KK tower \eqref{tower-operator}. Under the global $SU(2)\times SU(2)$ symmetry, this operator transforms in the representation $(\nicefrac{1}{2},\nicefrac{1}{2})$, in agreement with the fact that the corresponding state arises from the $T^{1,1}$ harmonics $l=j=\nicefrac{1}{2}$. It remains to be seen whether a truncation to this chiral multiplet is consistent; however, such a truncation is suggested by the fact that it appears at the bottom of a KK tower. One can consistently truncate the Lagrangian \eqref{LL} with $\beta = 0$ to a single complex scalar. Setting $\zeta_i = (\tanh(\eta/2) e^{i \varphi},0)$, we find
\bea
 && - \tfrac12 (\partial\eta )^{2}- \tfrac12 \sinh^2\eta (\partial \varphi - \tfrac12 \sqrt{3} A)^{2} \notag \\
&& - \tfrac{3}{8} \big(\cosh^2\eta -12 \cosh\eta -21
\big) . \qquad \label{beta=0}
\eea
The potential has  two critical points: a relative maximum at the origin, $\eta=0$, where it takes the value $-12$, and an absolute minimum at $\cosh(\eta)=6$. It increases exponentially for larger values of $\eta$ .

For the reasons explained above, if  a consistent trunctation  exists describing the dual dynamics of the chiral operator (\ref{operator}) with lowest conformal dimension in KW theory, under plausible assumptions the corresponding gravity model must be given by the $\beta=0$ Lagrangian. 
We now turn to the thermodynamics of the two different models arising on $T^{1,1}$. In holographic superconductors,  several instabilities can occur when the theory is considered at finite density. An instability leads to the condensation of a charged scalar operator below some critical temperature, which depends on the conformal dimension and the charge of the operator. 
In particular, one naturally expects that the phase transition that spontaneously breaks the $R$ symmetry is caused by the chiral primary operator 
of lowest dimension. Ref.~\cite{Gubser:2009qm} explicitly shows the  condensation of the chiral operator \eqref{tower1} with $\beta = 1$ and $\Delta = 3$. It also discusses the operator \eqref{operator} with $\beta = 0$ and $\Delta = \nicefrac32$ but, lacking a gravity model for this operator, they could only do so in the linearized approximation. Such an approximation is not conclusive  to determine if an operator actually condenses. In particular, one must check that the free energy of the corresponding hairy black hole is lower than the free energy of any other black hole in the theory
with the appropriate boundary behavior to contribute to the same thermal ensemble, including the \RN (RN) black hole describing the uncondensed phase.
Furthermore, the critical temperature determined by a linearized approximation will not be the actual critical temperature if the phase transition is 1st order.
Moreover, in some examples, the hairy black hole solution exists at  temperatures {\it above} a certain critical temperature.
This phenomenon was refered to as retrograde condensation in \cite{Aprile:2011uq} and has an analog in real condensed matter systems.
To exclude these possibilities one needs to go beyond the linearized approximation. Such a higher-order analysis  requires  the knowledge of the full scalar potential, which is provided by our $\beta =0$ model.

In $AdS$/CFT,  the  field theory at finite temperature and   finite chemical potential is described in terms of charged black holes with 
regular horizons.
The ansatz for the metric and the Maxwell gauge potential is
\be 
 ds^2=-g(r)e^{-\chi(r)}dt^2+\frac{dr^2}{g(r)}+ r^2 d \vec{x}^2, \ \ A=\Phi(r)dt  . \notag
\label{AnsatzEQM}
\ee
The generic solution has the following asymptotic,
\bea
e^{-\chi}g & = & e^{-\chi_{\infty}} \Big( r^2 -\frac{M}{r^2} + \ldots \Big) ,\ \ 
\Phi	   =  \mu-\frac{\rho}{r^2} + \ldots\ ,\nonumber\\
\eta  &=&  \frac{\mathcal{O}^{(1)}}{r^{4-\Delta}}+\frac{\mathcal{O}^{(2)}}{r^{\Delta}} + \ldots\ . \notag
\eea
where $\Delta$ is the greatest root of the equation $\Delta (\Delta - 4)=m^2$.  
The parameters $\mu$ and $\rho $ represents chemical potential and  charge density of the dual field theory system.
The coefficient $\mathcal{O}^{(1)}$ is interpreted as a source coupled to the operator, whose vacuum expectation value (VEV) is represented by the coefficient $\mathcal{O}^{(2)}$. 
For the condensed phase to arise as SSB of the global $U(1)$ current in the field theory, one must impose that the source coefficient $\mathcal{O}^{(1)}$ vanishes on the gravity solution. However, for the scalars with $m^2= - \nicefrac{15}{4}$,  the mode with fall-off $ 1/r^{4-\Delta}$ is normalizable as well. Therefore it is possible to adopt an alternative quantization where the role of sources and VEVs is exchanged \cite{Klebanov:1999tb}. Then one may choose to set $\mathcal{O}^{(2)}=0$ and look for SSB produced by the condensation of the operator with dimension $4-\Delta$, whose VEV will be given by $\mathcal{O}^{(1)}$.  In what follows we will consider both standard and alternative quantisation schemes, and investigate the thermodynamic competition between condensates with dimensions $3$, $\nicefrac32$ and $\nicefrac52$ for $T^{1,1}$, or in other words between the models \eqref{dallag} and \eqref{beta=0}.

\begin{figure}[t]
  \centering
  \subfiguretopcaptrue
  \subfigure[]{\includegraphics[scale=.5]{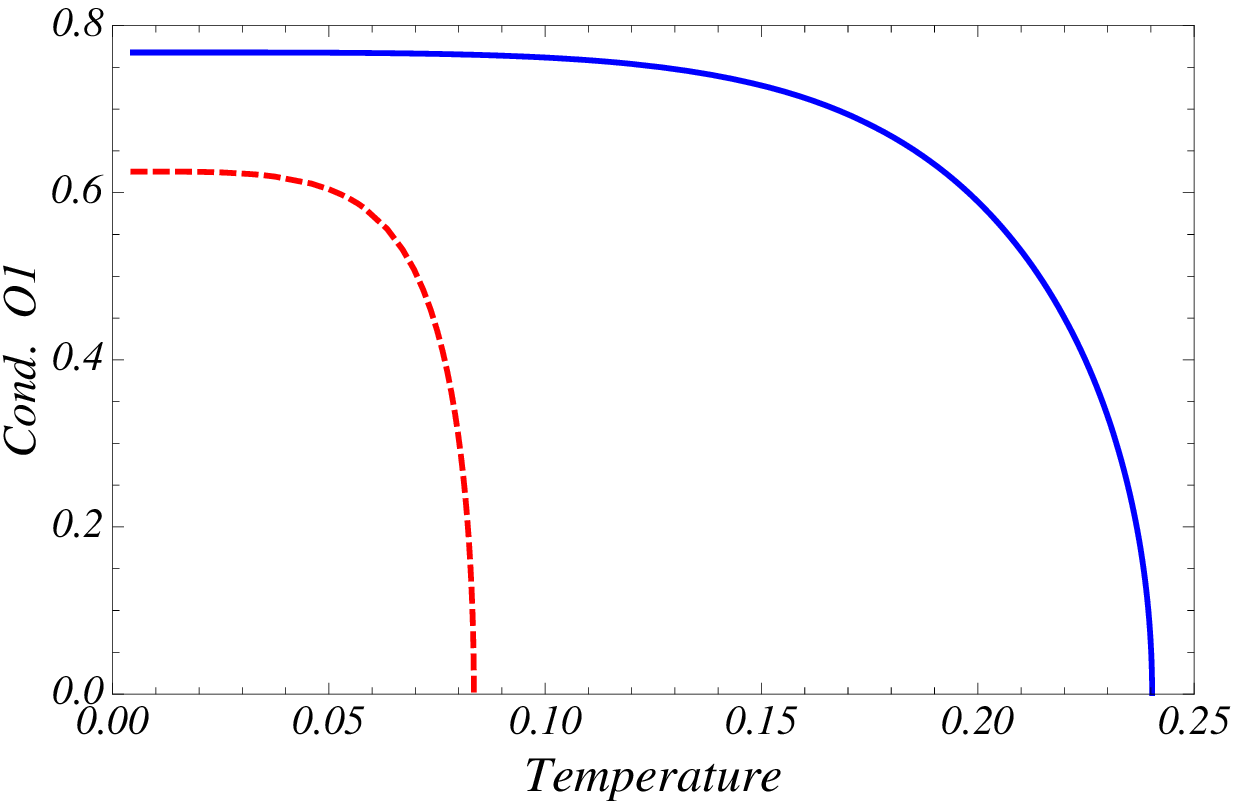}} \\ 
  \subfigure[]{\includegraphics[scale=.5]{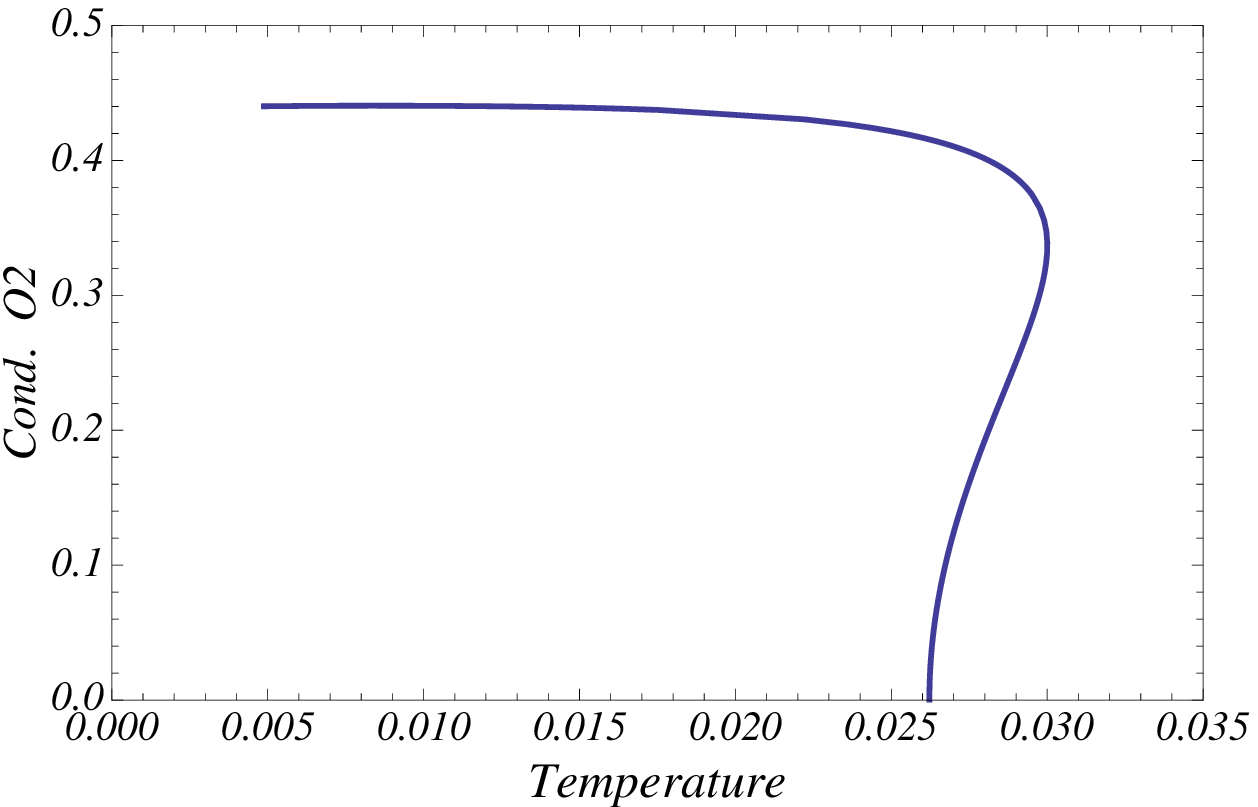}} 
  \caption{\it \small{The condensates $\langle O \rangle$ for the operators of conformal dimensions $\Delta= \nicefrac32$  and $\Delta=3$ (solid blue and dashed red lines in top fig.) and $\Delta= \nicefrac52$ (bottom fig.).}}
	\label{Figura2}
\vspace{-1.1cm}
\end{figure}

Consider first the $\beta =1$ Lagrangian (\ref{dallag}) describing the $m^2=-3$ complex scalar of \cite{Gubser:2009qm} coupled to the complex dilaton field 
$\tau $. 
The thermodynamics of $\eta$  at   {\it constant} dilaton was studied in \cite{Gubser:2009qm} as a model for holographic superconductors coming from  superstring theory. The dilaton is  coupled to the $m^2 =-3$ scalar in a non-trivial way. It is interesting to see if there are more general black holes with non-trivial dilaton.  In this model the equation of motion for the dilaton  can be integrated to
\bea
  \phi'(r)=c\frac{e^{\chi/2}}{r^3 g}\textrm{sech} \left(\tfrac12 \eta \right)\ . \notag
\eea
Since $\chi(r)$ is finite at the horizon, and $g(r_h)=0$, regularity of $\phi $ at the horizon requires that the integration constant $c$ vanishes. A similar analysis shows that $C_0$ must also be constant. 
This leads to the constant dilaton solution with non-trivial $\eta$, which was already described in \cite{Gubser:2009qm}. 
The corresponding condensate is reproduced by the  dashed red line in fig.~\ref{Figura2}.$1$. 

\begin{figure}[t]
  \centering
  \subfiguretopcaptrue
  \subfigure[]{\includegraphics[scale=.4]{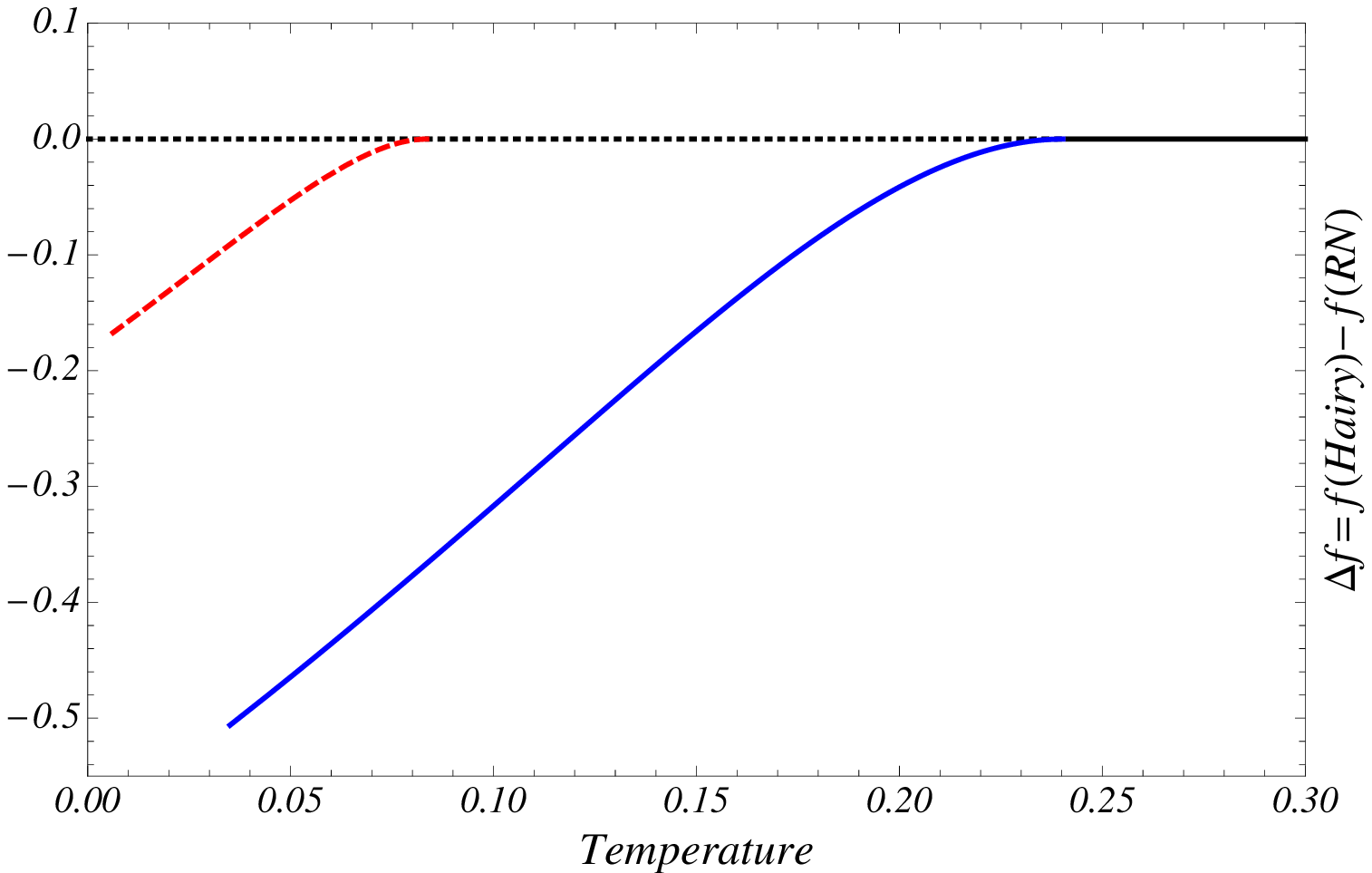}} \\ 
  \subfigure[]{\includegraphics[scale=.4]{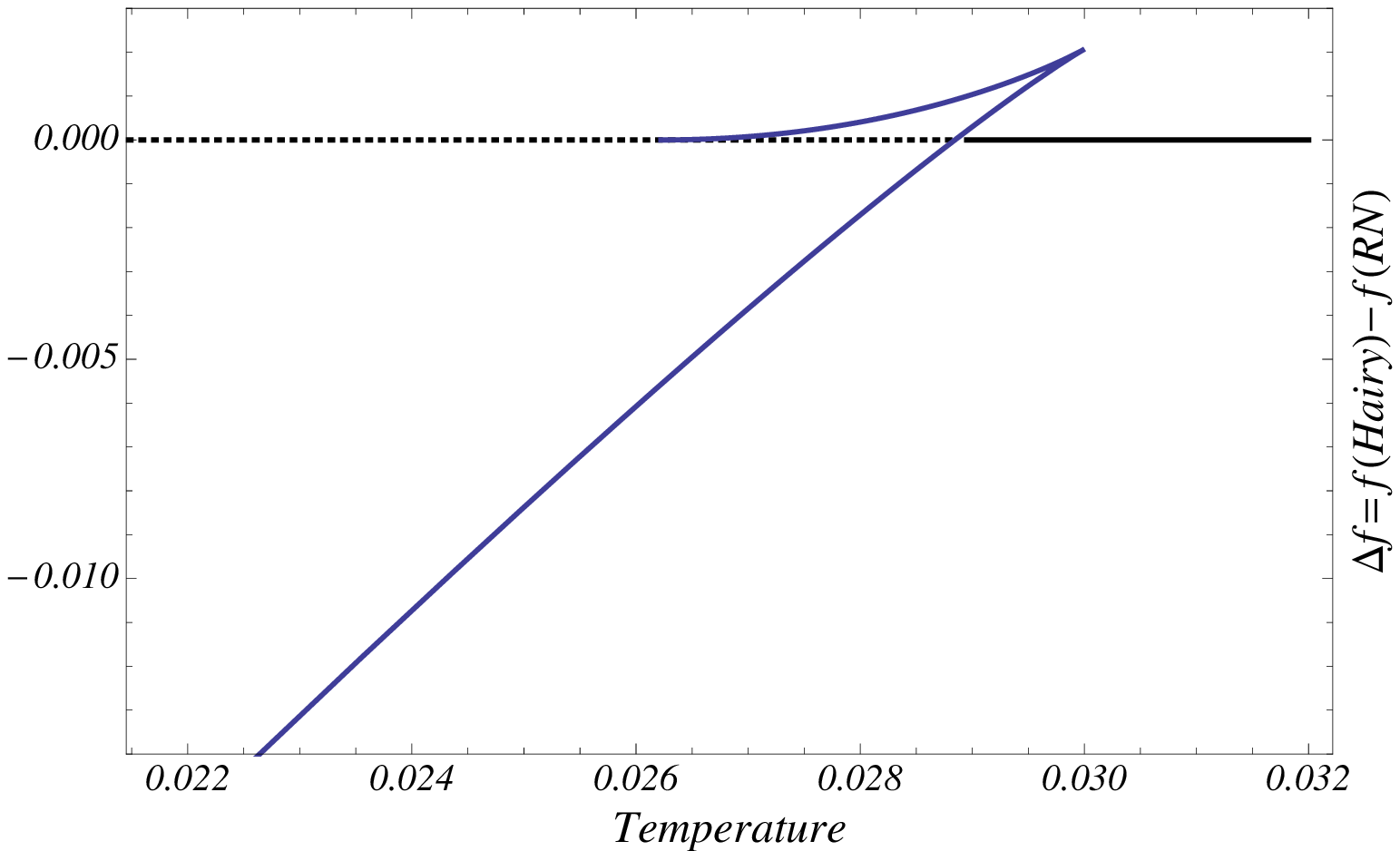}} 
  \caption{\it \small{The free energy relative to RN for the operators of conformal dimensions $\Delta = \nicefrac32$ and $\Delta = 3$ (solid blue and dashed red lines in top fig.) and $\Delta=\nicefrac52$ (bottom fig.).}}
	\label{Figura3}
\vspace{-.8cm}
\end{figure}

Next we turn to the $\beta = 0$ model. We have checked that  generic boundary conditions  at the horizon
lead either to a solution with $\{\zeta_+,\zeta_-\}=\{\zeta_+(r),0\} $ or $\{0,\zeta_-(r)\} $.
On the other hand, enforcing $\zeta_+=\zeta_-$ and solving the equations leads to retrograde condensation with a free energy
that is higher than the RN black hole. Thus consider the case $\zeta_-=0$,  described by \eqref{beta=0}.  
First opting for alternative quantisation, we seek a non trivial scalar profile for $\eta$ with asymptotics $O^{(2)}=0$.
Solving numerically the equations including back reaction, we find the family of solutions represented by the solid blue line in fig.~\ref{Figura2}$.1$.
It describes the VEV of the operator \eqref{operator} with $\Delta = \nicefrac32$. We note that this operator condenses at higher temperatures than  the dimension $\Delta=3$ operator which is dual to the $m^2=-3$ scalar in the $\beta =1$ model. We also find that the phase transition is 2nd order.
Figure \ref{Figura3}.$1$ compares the free energy of the black holes coming from the $\beta=1 $ and $\beta = 0$ scalar hair. We see that the thermodynamics is dominated  by  the phase in which the operator \eqref{operator} condenses for $T<T_c$, as this phase has the lowest free energy.

Finally,  fig.~\ref{Figura2}.$2$  shows the order parameter for the dimension $\Delta=\nicefrac52$ operator. In this case the standard quantization scheme is adopted. As expected  the solution exists at temperatures far below  the critical temperatures of the $\Delta =\nicefrac32$ and $\Delta=3$  operators. Therefore it does not contribute significantly to the thermodynamics.
The phase transition is 1st order. In this case the critical temperature is defined by the temperature at which the free energy becomes lower
than that of the RN black hole, see fig.~\ref{Figura3}.$2$. The phase transition is discontinuous because  the condensate has a jump  at $T_c$
from zero to a non-zero value. Strikingly, the general picture for the $\beta=0$ model is similar to the analogous 4D model of \cite{Bobev:2011rv}. 

To summarize, we have explicitly constructed a Lagrangian for the five-dimensional ${\cal N}=2$ gauged supergravity coupled to an  $SU(2,1)/U(2)$ hypermultiplet. The resulting model is uniquely determined by a single parameter $\beta$ representing the mixing between the $U(1)$ generators of $SU(2)$ and $U(1)$. When $\beta =1$, it describes two complex scalars of $m^2=-3$ and $m^2=0$. In this case it exactly coincides with the Lagrangian of \cite{Gubser:2009qm}, with the extension that incorporates the complex dilaton found in \cite{Cassani:2010uw, Liu:2010sa, Gauntlett:2010vu, Skenderis:2010vz}. This match involves a non-trivial scalar potential and non-trivial couplings, and should not come as a surprise as there is no other possible model for an $SU(2,1)/U(2)$ hypermultiplet with such masses.

Similarly, the same uniqueness property of the Lagrangian strongly indicates that the model with $\beta =0$ indeed must describe the two complex scalar fields 
of masses $m^2=- \nicefrac{15}{4}$ which are dual to the operator of lowest dimension $\Delta= \nicefrac32$ in the KW superconformal theory. We have explicitly demonstrated that this mode dominates the thermodynamics at low temperatures. It would be extremely interesting to see if the $\beta=0$ model represents a consistent truncation of type IIB supergravity. While the scalar fields have non-trivial KK quantum numbers $(\nicefrac{1}{2},\nicefrac{1}{2})$, they are the lowest states in the KK spectrum, which suggests that the truncation might nevertheless be consistent. Proving the latter may require an explicit construction of a type IIB ansatz that reproduces the same  equations of motion.

\smallskip

J.G.R.~would like to thank James Liu for a useful discussion. F.A.~would like to thank the hospitality of Perimeter Institute for Theoretical Physics. 
This research is supported by the following grants: MEC FPU  AP2008-04553, MCYT  FPA 2010-20807,
 a VIDI grant from NWO and CONACyT grants No.306769 and 
2009SGR502.

\providecommand{\href}[2]{#2}\begingroup\raggedright\endgroup

\end{document}